\documentclass{article}
\usepackage{spconf,amsmath,graphicx}
\usepackage{soul}
\usepackage{hyperref}
\usepackage{siunitx}
\usepackage{multirow}
\usepackage{multicol}
\usepackage{dcolumn}
\usepackage{tikz}
\usepackage{pgfplots}
\usepackage{diagbox}
\usepackage{amssymb}

\usepgfplotslibrary{statistics}
\pgfplotsset{compat=newest}

\makeatletter
  \g@addto@macro{\UrlBreaks}{\UrlOrds}
\makeatother

\definecolor{col1}{RGB}{211,47,47}
\definecolor{col2}{RGB}{123,31,162}
\definecolor{col3}{RGB}{0,151,167}
\definecolor{col4}{RGB}{46,125,50}

\makeatletter
\pgfplotsset{
  grid style={dotted, gray},
  boxplot/lower notch/.initial=\pgfplotsboxplotvalue{median},
  boxplot/upper notch/.initial=\pgfplotsboxplotvalue{median},
  boxplot/notch width/.initial=0.5,
  boxplot/draw/box/.code={%
    \draw[/pgfplots/boxplot/every box/.try]
      (boxplot box cs:\pgfplotsboxplotvalue{lower quartile},0)
      -- (boxplot box cs:\pgfplotsboxplotvalue{lower notch},0)
      -- (boxplot box cs:\pgfplotsboxplotvalue{median},0.5-\pgfplotsboxplotvalue{notch width}/2)
      -- (boxplot box cs:\pgfplotsboxplotvalue{upper notch},0)
      -- (boxplot box cs:\pgfplotsboxplotvalue{upper quartile},0)
      -- (boxplot box cs:\pgfplotsboxplotvalue{upper quartile},1)
      -- (boxplot box cs:\pgfplotsboxplotvalue{upper notch},1)
      -- (boxplot box cs:\pgfplotsboxplotvalue{median},0.5+\pgfplotsboxplotvalue{notch width}/2)
      -- (boxplot box cs:\pgfplotsboxplotvalue{lower notch},1)
      -- (boxplot box cs:\pgfplotsboxplotvalue{lower quartile},1)
      -- cycle
    ;
  },
  boxplot/draw/median/.code={%
    \draw[/pgfplots/boxplot/every median/.try]
        (boxplot box cs:\pgfplotsboxplotvalue{median},0.5-\pgfplotsboxplotvalue{notch width}/2)
        --
        (boxplot box cs:\pgfplotsboxplotvalue{median},0.5+\pgfplotsboxplotvalue{notch width}/2)
    ;
  }
}

\usepackage[colorinlistoftodos,prependcaption,textsize=tiny]{todonotes}

\def\R{{\mathbb R}}
\def\mb{\boldsymbol}
\def\eg{e.g.\ }
\def\ie{i.e.\ }
\def\wrt{w.r.t.\ }
\def\etal{et al.\ }
\def\Fig{Fig.\ }
\def\dSTOI{$\Delta\text{STOI}$}
\def\dSNRt{\Delta\text{SNR}_t}

\title{CLCNet: Deep learning-based noise reduction for hearing aids using complex linear coding}
\name{H. Schr\"oter$^{1}$, T. Rosenkranz$^{2}$, A. N. Escalante-B$^{2}$, M. Aubreville$^{1, \star}$, A.
Maier$^{1}$} \address{ $^1$ Friedrich-Alexander-Universit\"at Erlangen-N\"urnberg, Pattern
Recognition Lab \\ $^2$ Sivantos GmbH, Research and development, Erlangen, Germany \\
\texttt{hendrik.m.schroeter@fau.de}\\  \thanks{$\star$ M. Aubreville is also affiliated with 
Sivantos GmbH, Erlangen, Germany.}}
\begin{document}
%
\maketitle
\begin{abstract}
  Noise reduction is an important part of modern hearing aids and is included in most commercially
  available devices. Deep learning-based state-of-the-art algorithms, however, either do not
  consider real-time and frequency resolution constrains or result in poor quality under very noisy
  conditions.

  To improve monaural speech enhancement in noisy environments, we propose CLCNet, a framework based
  on complex valued linear coding. First, we define complex linear coding (CLC) motivated by linear
  predictive coding (LPC) that is applied in the complex frequency domain. Second, we propose a
  framework that incorporates complex spectrogram input and coefficient output. Third, we define a
  parametric normalization for complex valued spectrograms that complies with low-latency and
  on-line processing.

  Our CLCNet was evaluated on a mixture of the EUROM database and a real-world noise dataset
  recorded with hearing aids and compared to traditional real-valued Wiener-Filter gains.

\end{abstract}
\section{Introduction}
\label{sec:intro}

  Noise reduction is an emerging field in speech applications and signal processing. Especially in
  the context of an aging society with an increase in spread of hearing loss, noise reduction
  becomes a fundamental feature for the hearing-impaired.

  Advances in deep learning recently improved the performance of noise reduction algorithms
  \cite{aubreville2018deep, erdogan2015phase, kumar2016speech}. It is common practice to transform
  the noisy time-domain signal into a time-frequency representation, for instance using a short-time
  Fourier transform (STFT). Usually, only the magnitude of the complex valued spectrogram
  \cite{aubreville2018deep,tan2019real} is used for noise reduction. Recent publications though,
  also focus on incorporating phase information in the reconstructed output by using the phase
  \cite{zheng2019phase} or the raw complex valued spectrogram \cite{tan2019complex} as input. Most
  approaches, however, work with high frequency resolution (high-res) spectrograms
  \cite{williamson2016complex, tan2019complex}. This exceedingly simplifies the noise reduction
  process, since a single scalar value per frequency bin is sufficient to attenuate a narrow
  frequency range. A low-res spectrogram on the other hand cannot resolve speech harmonics that have
  a typical distance of \SIrange{50}{400}{\Hz}. Hence, a scalar factor is not able to reduce the
  noise between the harmonics while persevering the speech. Although a single complex factor could
  be used to perfectly reconstruct the clean signal via the complex ideal ratio mask (cIRM), it is
  hard to learn \cite{williamson2016complex, tan2019complex}. Due to a superposition of multiple,
  quasi static signals within one frequency bin, not only the phase changes over time but also the
  magnitude as a result of cancellation. Thus, low-res spectrograms limit the effectiveness of 
  standard complex valued processing methods that mainly bring phase improvement 
  \cite{zheng2019phase,roux2019phasebook}.

  While other work uses the whole signal in an off-line processing fashion as input for the noise
  reduction \cite{williamson2018monaural, wang2018end, zhao2018convolutional}, our work requires
  real-time capabilities. Both high-res spectrograms and off-line processing are not feasible for
  hearing aid applications, where the overall latency is a very important property. The
  superposition of both signals results in a clearly audible comb filter effect that is generating a
  tonal sound impression especially of background noises.  Therefore, an overall latency of
  \SI{10}{\ms} is typically the maximum of what is acceptable \cite{agnew2000just}.  Since higher
  frequency resolutions introduce bigger delays \cite{bauml2008uniform}, noise reduction for hearing
  aids needs to be performed on low-res spectrograms. This, and on-line processing constraints are
  not considered in any state-of-the-art (SOTA) algorithms.

  To overcome these limitations, this study proposes a framework motivated by LPC. Due to the low
  resolution, one frequency band can contain multiple harmonics. This results in a superposition of
  multiple complex valued periodic signals for each frequency band. LPC is able to perfectly model
  a superposition of multiple sinusoidals given enough coefficients. Due to this property, LPC finds
  a use case in speech coding and synthesis \cite{valin2019lpcnet}. Yet, it is often only applied on
  time-domain signals as a post-processing step. Instead, we propose a complex valued linear
  combination of the model output and the noisy spectrogram. We can show that this outperforms
  previous approaches like real valued Wiener-Filter (WF) masking on low-res spectrograms.

\section{Related work}
\label{sec:related_work}

  Removing unwanted environmental background noise in speech signals is a common step in speech
  processing applications. Complex valued neural networks as well as phase estimation have been of
  great interest in speech enhancement lately, since the perceptual audio quality has been reported
  to be improved significantly \cite{zheng2019phase, williamson2016complex, tan2019complex,
  wang2018end}.

  One approach is to estimate magnitude and phase or a phase representation either directly
  \cite{zheng2019phase} or use the estimated magnitude and noisy phase to predict the clean phase
  \cite{roux2019phasebook}. Estimating the clean phase directly, however, is quite hard, because of
  its spontaneous, random-like nature. Zheng \etal \cite{zheng2019phase} jointly estimated the
  magnitude spectrogram and a phase representation based on the time derivative of the phase. Le
  Roux \etal \cite{roux2019phasebook} estimated a magnitude mask and clean phase using a
  cookbook-based method. Reducing the phase estimate from a continuous space to discrete cookbook
  values reduces the search space and allows to use output activations that are designed to
  represent the cookbook values, like a convex softmax.

  Other work focuses on estimating a complex valued mask. Williamson \etal
  \cite{williamson2016complex} used a complex ratio mask (CRM) to reconstruct a clean speech
  spectrogram. Yet the network did not use complex valued input features, but used traditional real
  valued features such as MFCCs as input. Tan \etal tried to directly estimate the complex valued
  spectrogram \cite{tan2019complex} using a linear output layer. This, however, might not be very
  robust, since the network is allowed to output any value.

  None of those SOTA algorithms, however, fulfill the latency and thus low-res spectrogram
  requirements. Even Tan \etal \cite{tan2019real}, who propose a convolutional recurrent network for
  real-time speech enhancement do not specify overall latency. However, the windowing used in their
  approach for the frequency transform results in \SI{20}{\ms} of delay. Furthermore, it was only
  evaluated on synthetic noise.

  Only Aubreville \etal \cite{aubreville2018deep} fully respects those constraints. Their approach
  uses Wiener-Filter (WF) gains to reduce unwanted environmental noise. The real-valued WF gains can
  only modify the magnitude of the enhancement spectrogram. This, however, performs poor on low
  signal to noise ratios (SNRs) since the WF is restricted by frequency resolution, resulting in
  phase distortions that decrease perceptual quality.

\vspace{-0.5em}
\section{Complex Linear Coding}
\label{sec:methods}

  In this section, we will describe details of our approach and provide a theoretical motivation.
  Starting with complex valued LPC, we derive a more general noise reduction framework based on a
  complex valued linear combination. Finally, we introduce a phase-aware, parametric normalization
  method for complex spectrograms. We provide details of our implementation, which is qualified to
  be embedded into a hearing aid signal processing chain, and the used database.

\subsection{Linear Predictive Coding}
\label{ssec:lpc}

  Linear predictive coding is known to be a suitable model of the vocal tract response
  \cite{makhoul1975linear} and is still used in SOTA approaches for speech coding and synthesis
  \cite{valin2019lpcnet}. Given a signal $x_k$ at sample $k$, LPC can be described as the linear
  combination
  \begin{equation}
    \hat{x}_k = \sum_{i=1}^N a_i x_{k-i} \text{\ ,}
    \label{eq:LPC}
  \end{equation}
  where $a_i$ are the LPC coefficients of order $N$ and $\hat{x}_k$ is the predicted sample at
  position $k$. Let $d_k$ be the prediction error
  \begin{equation}
    \vspace{-0.2em}
    d_k = x_k - \hat{x}_k = x_k - \sum_{i=1}^N a_i x_{k-i} \text{\ .}
  \end{equation}
  The optimal coefficients $a_i$ can then be found by minimizing the expectation value $E\{d_k^2\}$.
  This results in $n$ equations that need to be solved. In practice, a solution can be found \eg via
  autocorrelation of $x$ and the Levinson-Durbin \cite{durbin1960fitting} algorithm. Higher order
  $n$ allow to model a higher number of superimposed frequencies in the signal $x$. For instance,
  Makhoul \etal \cite{makhoul1975linear} showed that $n=10$ coefficients are enough to sufficiently
  model the dominant frequencies of the vocal tract.

  Although LPC is often only applied to real-valued time-domain signals, it is equivalent for a
  complex valued signal. In the next section, we describe our proposed noise reduction framework
  using a general form of equation (\ref{eq:LPC}).

\subsection{Noise reduction via complex linear combination}
\label{ssec:clnr}
  \begin{figure}[tb]
    \centering
    \includegraphics[width=\linewidth,trim=1.2cm 0.1cm 1.0cm 0.5cm,clip]{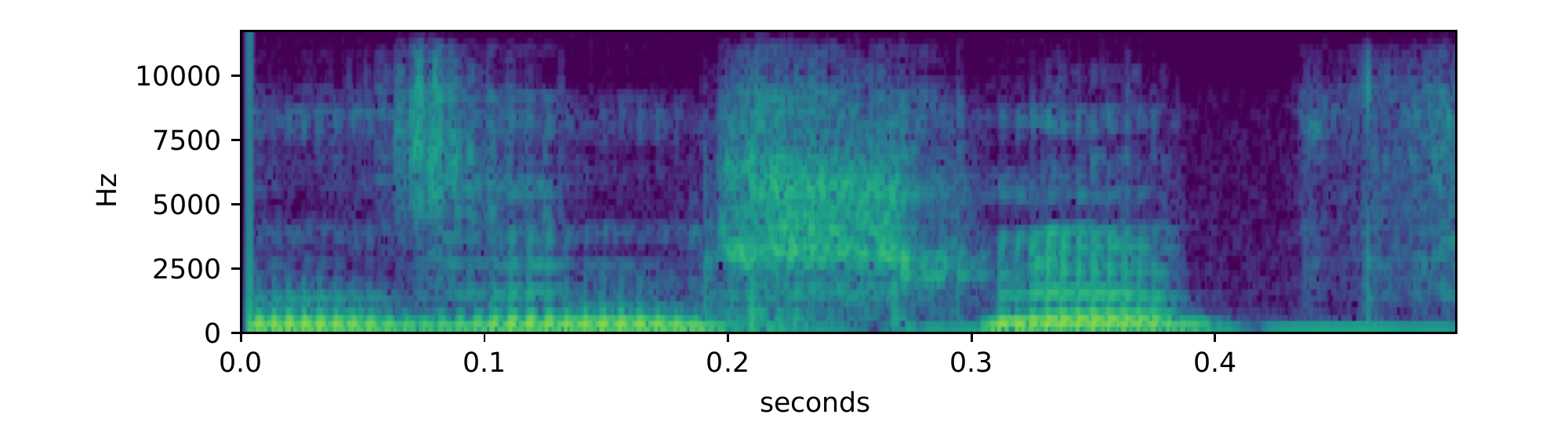}
    \vspace{-2.0em}
    \caption{A power spectrogram using the deployed filter bank from a clean speech sample of the
    train set. This filter bank is similar to a 96 point STFT. The speech-typical harmonics
    are not clearly visible due to the low frequency resolution.}
    \label{fig:clean_spec}
    \vspace{-1.0em}
  \end{figure}
  Spectrograms have a periodic structure if the underlying time domain signal is a superposition of
  sinusoidals and the frequency resolution of the spectrogram cannot resolve the different
  frequencies in the signal. This is caused by cancellation within each frequency band.  Since those
  spectrograms have wide frequency bands, multiple harmonics of human speech can lie within a
  frequency band. Due to the overlap between two successive windows during STFT, the phase of the
  different frequencies in the time-domain signals changes with different rotation speed. This may
  lead to partial cancellation, \ie the magnitude of the superposition of those frequencies may
  decrease. This effect can be observed in \Fig \ref{fig:clean_spec}. Here, a single frequency band
  is approximately \SI{500}{\Hz} wide. Assuming a minimal human fundamental frequency $f_0=100$, up
  to 5 harmonic oscillations can be captured within a band.

  To enhance a noisy spectrogram, a naive approach would be to calculate the LPC coefficients of the
  ideal clean speech via Levinson-Durbin and apply it to the noisy spectrogram.  A deep
  learning-based model would learn the mapping from a noisy spectrogram to the ideal LPC
  coefficients. This however, does not work very well. First of all, the coefficients computed from
  the clean spectrogram are only meaningful for time-frequency (TF) bins that include harmonic parts
  of speech. For TF bins without speech, the LPC coefficients will not enhance the resulting
  spectrogram. Furthermore, the ``ideal'' LPC coefficients only slightly reduce white noise and do
  not enhance the noisy spectrogram \wrt any metric, \ie amplitude (IAM) or energy (WF).

  Instead we propose a complex linear coding (CLC) framework. Since we know that LPC modeled by a
  complex linear combination works well for for harmonic signals like speech, we embed the linear
  combination as a known operator \cite{maier2019learning} in the network. Given a noisy
  spectrogram, the model predicts complex valued coefficient that are applied to the noisy
  spectrogram again. Thus, CLC will output an enhanced spectrogram that can be transformed into
  time-domain. The loss can then be computed in either time or frequency domain.
 
  In contrast to LPC, for CLC we can use information of the current and even future frames resulting
  in a more general form of the linear combination in (\ref{eq:LPC}):
  \begin{equation}
    \vspace{-0.2em}
    \mb{\hat{S}}(k, f) = \sum_{i=0}^{N} \mb{A}(k, i, f) \cdot \mb{X}(k - i + l, f)\text{\ ,}
    \label{eq:CLC}
  \end{equation}
  were $l$ is an offset and $N$ the order of the linear combination. $\mb{A}(k,i,f)$ are the
  output coefficients with $i=0,\dots,N$ for each time-index $k$ and frequency-index $f$. For
  $l=-1$, this is equivalent to LPC. Note that $\mb{S}$, $\mb{A}$ and $\mb{X}$ are
  complex, thus the multiplication needs to be complex valued.

  As described above, one frequency bin can include up to 5 speech harmonics. Thus, we chose a CLC
  order of $N=5$ for our noise reduction framework.

\subsection{Parametric unit norm normalization}
\label{ssec:normalization}

  Normalization is an essential part of most deep learning pipelines which helps for robustness and
  generalization. In speech processing applications, most normalization methods are performed on
  power-spectrogram level which is not applicable for complex valued input. Instead, we propose a
  bin-wise and phase-sensitive normalization scheme based on the unit norm. Given a filter bank
  representation $\mb{X}(k,f)$, the signal is normalized as:
  \begin{equation}
    \vspace{-0.2em}
    \mb{X}_{\text{norm}}(k, f) = \frac{\mb{X}(k, f)}{\mu_{k, f}} \cdot \gamma_{f}\text{ ,}
    \vspace{-0.2em}
    \label{eq:norm}
  \end{equation}
  where $\mu$ is the mean of $|\mb{X}|$ and $\gamma_f \in \R$ a learnable parameter for each
  frequency bin. $\mu$ can be computed in a real-time capable fashion like a exponential moving
  average or a window-based approach. Since the complex valued input is only multiplied with scalar
  values, the phase of $\mb{X}$ does not change.

\subsection{Hearing aid signal processing chain}
\label{ssec:processing_chain}

  Instead of a usual STFT, our signal processing chain employs a standard uniform polyphase filter
  bank for hearing aids \cite{bauml2008uniform}. In particular, a 48-frequency-bin analysis filter bank
  transforms the time-domain signal of clean speech $s$, noise $n$ and noisy mixture $m$ into
  the representations $\mb{S}(k,f)$, $\mb{N}(k,f)$, $\mb{M}(k,f)$.

  We directly feed the complex valued filter bank representations into a parametric channel-wise
  normalization to enhance the weaker frequency bands. After the noise reduction step via complex
  linear combination (\ref{eq:CLC}), the enhanced filter bank representation $\mb{\hat{S}}(k,f)$
  gets synthesized again.

\section{Experiments}
\label{sec:experiments}
\subsection{Dataset and Implementation Details}
\label{ssec:impl}

  The clean speech corpus contains 260 German sentences from 52 speakers from the EUROM database
  \cite{chan1995eurom} upsampled to \SI{24}{\kHz}. We furthermore used 49 real-world noise signals
  including non-stationary signals to generate noisy mixtures. The noise signals were recorded in
  various places in Europe using hearing aid microphones in a receiver in a canal-type hearing
  instrument shell (Signia Pure 312, Sivantos GmbH, Erlangen, Germany) and calibrated recording
  equipment at a sampling rate of \SI{24}{\kHz}. The mixtures were created with up to $4$ noises at
  various SNRs of $\{-100, -5, 0, 5, 10, 20\}\ \si{\dB}$ and level offsets $\{-6,0,6\}\ \si{\dB}$
  sampled online with an uniform distribution. The dataset was split on original signal level into
  train, validation and test sets, where validation set was used for model selection and test set to
  report the results.

  \begin{figure}[b]
    \vspace{-1.0em}
    \includegraphics[width=\columnwidth, trim=0 20 0 5, clip]
      {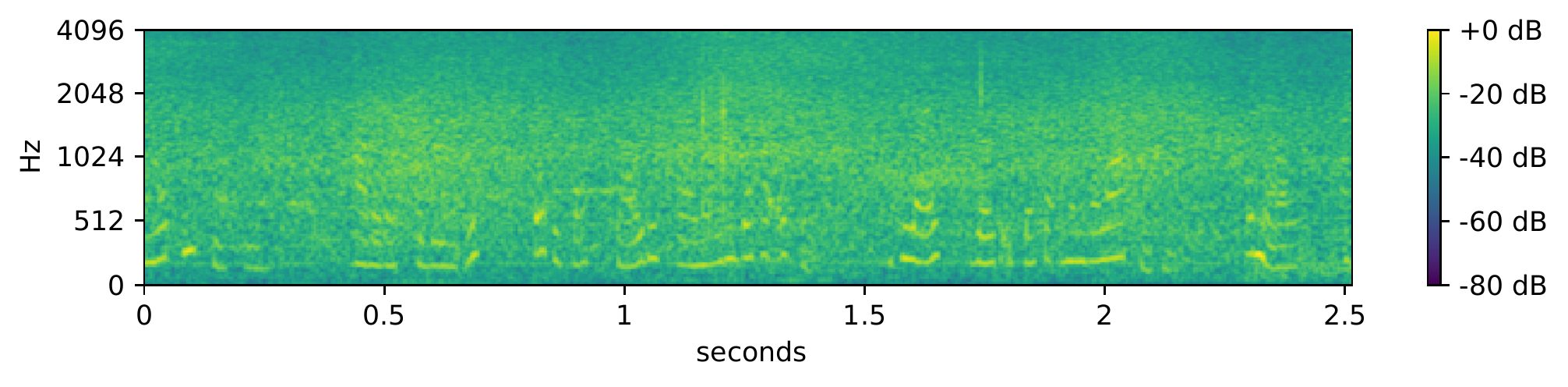}
    \includegraphics[width=\columnwidth]
      {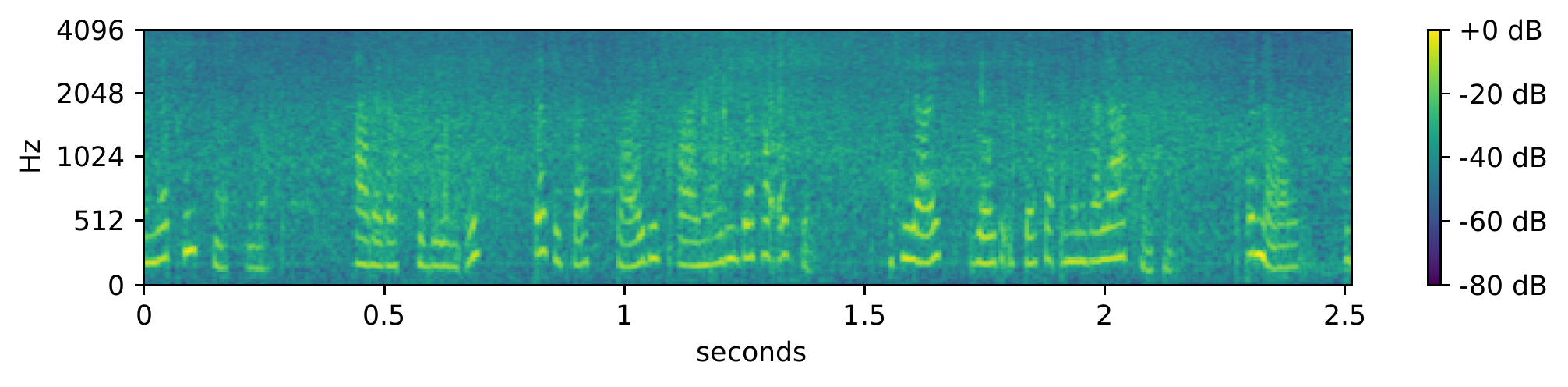}
    \vspace{-2.2em}
    \caption{High resolution Mel spectrograms of a noisy input and an enhanced output.}
    \label{fig:melspectrograms}
  \end{figure}

  Our framework was implemented in PyTorch \cite{paszke2017automatic} using a similar MLP-based
  architecture (3 fully connected layers with ReLU activations) and temporal context to
  \cite{aubreville2018deep}. Only the input and output layers were modified due to the complex
  filter bank representation input and coefficient output. Thus, we chose a tanh activation to also
  allow negative output. The complex input spectrogram is normalized given the temporal context of
  $\tau=\tau_1+1+\tau_2$, where $\tau_1=\SI{200}{ms}$ is the look-back and $\tau_2=\SI{2}{ms}$ the
  lookahead context.  We used an Adam optimizer with an initial learning rate of $3\cdot10^{-4}$. The loss
  was computed in time-domain using a multi-objective loss of RMSE loss and scale-invariant signal
  distortion ratio (SI-SDR) \cite{roux2019sdr}. We found that the SI-SDR helped to enhance higher
  harmonics of the original speech, while the RMSE loss penalized the RMS difference of the enhanced
  audio. The scale-dependent SDR did not converge in out experiments.

  The maximum attenuation of the noise reduction was limited to \SI{14}{\dB} similar to
  \cite{aubreville2018deep}, because in practice not all noise should be removed for hearing aid
  applications. Since the attenuation cannot be limited afterwards, the model was trained to only
  improve the SNR by \SI{14}{\dB}. Therefore, the clean speech was not used directly as target but
  rather a mixture with an $\dSNRt = \SI{14}{\dB}$ over the noisy input signal. \Fig
  \ref{fig:melspectrograms} shows an utterance of the test set with an SNR of \SI{-5}{\dB}, enhanced
  with an attenuation of $\dSNRt = \SI{14}{\dB}$. German and English audio samples are available at
  \cite{schroeter2019clc}.

\subsection{Objective Evaluation and Discussion}
\label{ssec:evaluation}

  The enhanced speech signals are evaluated using two objective metrics, namely the scale-invariant
  signal distortion ratio (SI-SDR) \cite{roux2019sdr} and the difference between noisy and enhanced
  signal \wrt the short-time objective intelligibility (STOI) \cite{taal2011algorithm}, denoted as
  \dSTOI. We evaluate different configurations of our framework and compare them with previous
  work \cite{aubreville2018deep}.

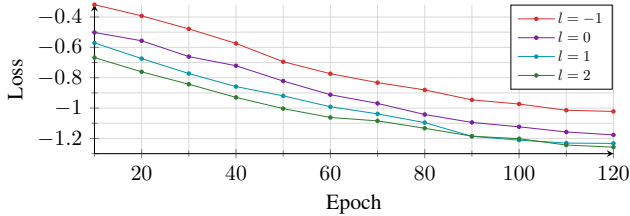
\begin{figure}[t]
    \centering
    \resizebox{\linewidth}{!} {
    \begin{tikzpicture}
      \begin{axis}[
        ymin=-1.3,
        xmax=120,
        width=10cm,
        height=4.0cm,
        grid=both,
        minor tick num=1,
        grid style={solid,gray!30!white},
        axis lines=left,
        xlabel={Epoch},
        ylabel={Loss},
        legend style={at={(0.9, 1.00)}, anchor=north, nodes={scale=0.7, transform shape}},
        legend cell align={left},
        legend image post style={mark=*},
        /pgf/number format/fixed,
        /pgf/number format/1000 sep=\thinspace]
        \addplot[color=col1, mark=*, thin, mark size=0.8pt]
          table [x=step, y=lc26_om1_smoothed, col sep=comma] {assets/validation_loss_lc.csv};
        \addplot[color=col2, mark=*, thin, mark size=0.8pt]
          table [x=step, y=lc25_o0_smoothed,  col sep=comma] {assets/validation_loss_lc.csv};
        \addplot[color=col3, mark=*, thin, mark size=0.8pt]
          table [x=step, y=lc19_o1_smoothed,  col sep=comma] {assets/validation_loss_lc.csv};
        \addplot[color=col4, mark=*, thin, mark size=0.8pt]
          table [x=step, y=lc20_o2_smoothed,  col sep=comma] {assets/validation_loss_lc.csv};
        \addlegendentry{$l=-1$};
        \addlegendentry{$l=0$};
        \addlegendentry{$l=1$};
        \addlegendentry{$l=2$};
      \end{axis}
    \end{tikzpicture}
    }
    \vspace{-2.2em}
    \caption{Validation loss for different offsets $l$. An offset of $l=-1$ corresponds to
      predicting the next frame like in LPC.}
    \label{fig:val_loss_offsets}
    \vspace{-1.5em}
  \end{figure}

  When comparing the loss for different offsets $l$ as shown in \Fig \ref{fig:val_loss_offsets}, we
  can see that the CLC-framework benefits from lookahead context. While an offset of $l=-1$ leads to a
  significant performance drop, the loss converges for a higher context as $l$ gets greater
  than 1. Thus we chose $l=1$ for further experiments.

  \begin{table}[b]
    \small
    \vspace{-2.0em}
    \caption{\small SI-SDR [\si{\dB}] test set performance for different SNRs.}
    \label{tab:res-sdr-14dB}
    \vspace{0.5em}
    \begin{minipage}[b]{\columnwidth}
      \centering
      \begin{tabular}{|l|l|r|r|r|r|r|}
        \hline
        Offset $l$ & 20 & 10 & 5 & 0 & -5 \\ \hline
         & \multicolumn{5}{c|}{$\dSNRt=\SI{14}{\dB}$} \\ \hline
        -1 & 21.64 & 15.73 & 12.27 & 8.06 & 3.24 \\ \hline
        0 & 21.18 & 15.69 & 12.21 & 8.07 & 3.14 \\ \hline
        1 & 22.27 & 16.19 & \textbf{12.70} & \textbf{8.53} & \textbf{3.63} \\ \hline
        2 & \textbf{22.74} & \textbf{16.28} & 12.58 & 8.18 & 3.20 \\ \hline \hline
         & \multicolumn{5}{c|}{$\dSNRt=\SI{100}{\dB}$} \\ \hline
        1 & 21.69 & \textbf{16.28} & \textbf{13.48} & \textbf{10.09} & \textbf{5.88} \\ \hline
      \end{tabular}
    \end{minipage}
  \end{table}

  The objective results in table \ref{tab:res-sdr-14dB} show the performance of different CLCNet
  configurations. We find that limiting the maximum attenuation $\dSNRt=\SI{14}{\dB}$ decreases
  SI-SDR especially for low SNRs like \SI{-5}{\dB}, which is an expected result. For higher SNRs,
  however, the difference is negligible.

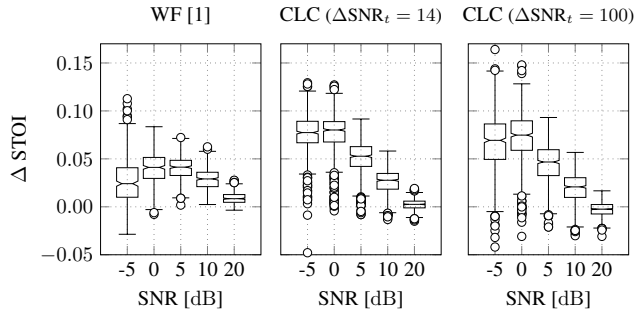
\begin{figure}[t]
    \vspace{-0.5em}
    \resizebox{\columnwidth}{!} {
      \begin{tikzpicture}
  \begin{axis}[
      xlabel={SNR [\si{\dB}]},
    title={WF \cite{aubreville2018deep}},
    xmin=0,
    xmax=6,
    ymin=-0.05,
    ymax=0.17,
    grid=major,
    xtick={1, 2, 3, 4, 5},
    xticklabels={-5, 0, 5, 10, 20},
    y tick label style={
        /pgf/number format/.cd,
            fixed,
            fixed zerofill,
            precision=2,
        /tikz/.cd
    },
    ylabel={$\Delta$ STOI},
    boxplot/draw direction=y,
    width=0.5\linewidth,
    height=0.6\linewidth,
    boxplotcolor/.style={color=#1,mark options={color=#1,fill=white}},
    ]

    \addplot+[boxplotcolor=black, mark=*,boxplot prepared={
            median=2.415000e-02, 
            upper quartile=4.075000e-02, 
            lower quartile=9.990000e-03, 
            upper whisker=8.679000e-02, 
            lower whisker=-2.864000e-02,
            lower notch=2.147118e-02,
            upper notch=2.682882e-02}
        ]
        coordinates {
          (1.0, 0.10198999999999991)
          (1.0, 0.094899999999999984)
          (1.0, 0.093560000000000088)
          (1.0, 0.10774000000000006)
          (1.0, 0.11273999999999995)
          (1.0, 0.090940000000000021)
          (1.0, 0.10018000000000005)};

    \addplot+[boxplotcolor=black, mark=*,boxplot prepared={
            median=4.111000e-02, 
            upper quartile=5.150000e-02, 
            lower quartile=2.963000e-02, 
            upper whisker=8.352000e-02, 
            lower whisker=-2.890000e-03,
            lower notch=3.919356e-02,
            upper notch=4.302644e-02}
        ]
        coordinates {
          (2.0, -0.0078399999999999581)
          (2.0, -0.0060700000000000198)};

    \addplot+[boxplotcolor=black, mark=*,boxplot prepared={
            median=4.136500e-02, 
            upper quartile=4.848250e-02, 
            lower quartile=3.267750e-02, 
            upper whisker=7.131000e-02, 
            lower whisker=9.350000e-03,
            lower notch=4.000725e-02,
            upper notch=4.272275e-02}
        ]
        coordinates {
          (3.0, 0.0018000000000000238)
          (3.0, 0.0089499999999999025)
          (3.0, 0.072249999999999925)
          (3.0, 0.008899999999999908)};

    \addplot+[boxplotcolor=black, mark=*,boxplot prepared={
            median=2.908000e-02, 
            upper quartile=3.603500e-02, 
            lower quartile=2.104000e-02, 
            upper whisker=5.785000e-02, 
            lower whisker=2.380000e-03,
            lower notch=2.782341e-02,
            upper notch=3.033659e-02}
        ]
        coordinates {
          (4.0, 0.060029999999999917)
          (4.0, 0.062549999999999994)};

    \addplot+[boxplotcolor=black, mark=*,boxplot prepared={
            median=8.460000e-03, 
            upper quartile=1.277500e-02, 
            lower quartile=4.697500e-03, 
            upper whisker=2.400000e-02, 
            lower whisker=-3.520000e-03,
            lower notch=7.757627e-03,
            upper notch=9.162373e-03}
        ]
        coordinates {
          (5.0, 0.02733999999999992)
          (5.0, 0.025349999999999984)
          (5.0, 0.027710000000000012)
          (5.0, 0.025419999999999998)
          (5.0, 0.025210000000000066)};
	\end{axis}
\end{tikzpicture}%
      \begin{tikzpicture}
  \begin{axis}[
    xlabel={SNR [\si{\dB}]},
    title={CLC \small($\Delta\text{SNR}_t=14$)},
    xmin=0,
    xmax=6,
    ymin=-0.05,
    ymax=0.17,
    grid=major,
    xtick={1, 2, 3, 4, 5},
    xticklabels={-5, 0, 5, 10, 20},
    yticklabels={},
    ytick={},
    boxplot/draw direction=y,
    width=0.5\linewidth,
    height=0.6\linewidth,
    boxplotcolor/.style={color=#1,mark options={color=#1,fill=white}},
    ]

    \addplot+[boxplotcolor=black, mark=*,boxplot prepared={
        median=0.07739350199699402,
        lower quartile=0.06678687036037445,
        upper quartile=0.08917789533734322,
        lower whisker=0.03418354690074921,
        upper whisker=0.12069659680128098,
        lower notch=0.07598509036417077,
        upper notch=0.07880191362981727}
        ]
        coordinates {
            (1.0, 0.0067000556737184525)
            (1.0, 0.02251128852367401)
            (1.0, 0.022183705121278763)
            (1.0, 0.033157844096422195)
            (1.0, 0.009943388402462006)
            (1.0, 0.02040339633822441)
            (1.0, 0.025244515389204025)
            (1.0, 0.008529748767614365)
            (1.0, 0.03024616278707981)
            (1.0, 0.030869266018271446)
            (1.0, -0.04789526015520096)
            (1.0, 0.016431940719485283)
            (1.0, 0.010164552368223667)
            (1.0, 0.030299849808216095)
            (1.0, -0.35018882155418396)
            (1.0, 0.011273916810750961)
            (1.0, 0.025466494262218475)
            (1.0, 0.003222779603675008)
            (1.0, 0.020436584949493408)
            (1.0, -0.008691953495144844)
            (1.0, 0.022415027022361755)
            (1.0, 0.016901377588510513)
            (1.0, 0.02718953788280487)
            (1.0, 0.014425297267735004)
            (1.0, 0.007341622840613127)
            (1.0, 0.1293582171201706)
            (1.0, 0.12541796267032623)
            (1.0, 0.12552407383918762)
            (1.0, 0.1280994862318039)
        };

    \addplot+[boxplotcolor=black, mark=*,boxplot prepared={
        median=0.08011863008141518,
        lower quartile=0.06759200058877468,
        upper quartile=0.0888450238853693,
        lower whisker=0.0360567532479763,
        upper whisker=0.1183302104473114,
        lower notch=0.07878287118348555,
        upper notch=0.0814543889793448}
        ]
        coordinates {
            (1.0, 0.0038204507436603308)
            (1.0, 0.03207719698548317)
            (1.0, 0.028888242319226265)
            (1.0, -0.001391748315654695)
            (1.0, 0.0032700386364012957)
            (1.0, 0.006512218154966831)
            (1.0, 0.005644732620567083)
            (1.0, 0.03194722160696983)
            (1.0, 0.015470935963094234)
            (1.0, 0.019633211195468903)
            (1.0, 0.01076115295290947)
            (1.0, -0.00421920046210289)
            (1.0, 0.014996717683970928)
            (1.0, 0.0006288464646786451)
            (1.0, 0.03340255841612816)
            (1.0, 0.008337673731148243)
            (1.0, 0.01003037765622139)
            (1.0, 0.004846806172281504)
            (1.0, 0.0111465472728014)
            (1.0, 0.01483076810836792)
            (1.0, 0.03551043942570686)
            (1.0, 0.0031828810460865498)
            (1.0, 0.029721679165959358)
            (1.0, 0.007423631381243467)
            (1.0, 0.024039169773459435)
            (1.0, 0.028834998607635498)
            (1.0, 0.019205031916499138)
            (1.0, 0.00809154286980629)
            (1.0, 0.023989610373973846)
            (1.0, 0.03120352327823639)
            (1.0, 0.0060493540950119495)
            (1.0, 0.001732243923470378)
            (1.0, 0.004183108918368816)
            (1.0, -0.18301858007907867)
            (1.0, 0.009896778501570225)
            (1.0, 0.026251748204231262)
            (1.0, 0.02052190899848938)
            (1.0, 0.026518439874053)
            (1.0, 0.0013057452160865068)
            (1.0, 0.022495418787002563)
            (1.0, 0.03277173638343811)
            (1.0, 0.12337944656610489)
            (1.0, 0.1270913928747177)
            (1.0, 0.12531687319278717)
            (1.0, 0.12579777836799622)
            (1.0, 0.12222670763731003)
        };

    \addplot+[boxplotcolor=black, mark=*,boxplot prepared={
        median=0.052767541259527206,
        lower quartile=0.042131612077355385,
        upper quartile=0.0627543218433857,
        lower whisker=0.01123390905559063,
        upper whisker=0.09157375246286392,
        lower notch=0.05147139775631987,
        upper notch=0.05406368476273454}
        ]
        coordinates {
            (1.0, 0.008134575560688972)
            (1.0, 0.010408321395516396)
            (1.0, 0.0029021527152508497)
            (1.0, 0.0035617805551737547)
            (1.0, 0.008500366471707821)
            (1.0, 0.009279155172407627)
            (1.0, 0.00899494532495737)
            (1.0, 0.004655554890632629)
            (1.0, 0.005433942191302776)
            (1.0, 0.006127957720309496)
            (1.0, 0.0011090284679085016)
            (1.0, 0.0015282257227227092)
            (1.0, 0.008671551011502743)
            (1.0, 0.002958917524665594)
            (1.0, -0.0036626849323511124)
            (1.0, 0.0057793306186795235)
            (1.0, 0.005764639470726252)
            (1.0, -0.007933743298053741)
            (1.0, 0.00882304459810257)
            (1.0, 0.009374020621180534)
            (1.0, 0.00814555399119854)
            (1.0, 0.007608294486999512)
            (1.0, 0.001523531274870038)
            (1.0, -0.008116117678582668)
            (1.0, -0.006054393015801907)
            (1.0, -0.004808390978723764)
            (1.0, 0.008243913762271404)
        };

    \addplot+[boxplotcolor=black, mark=*,boxplot prepared={
        median=0.027700698003172874,
        lower quartile=0.018394722137600183,
        upper quartile=0.03479685354977846,
        lower whisker=-0.0061427815817296505,
        upper whisker=0.05816137418150902,
        lower notch=0.026669819117233914,
        upper notch=0.028731576889111834}
        ]
        coordinates {
            (1.0, -0.008058559149503708)
            (1.0, -0.012065856717526913)
            (1.0, -0.010087513364851475)
            (1.0, -0.00858344230800867)
            (1.0, -0.007287899497896433)
            (1.0, -0.007194378413259983)
            (1.0, -0.009089947678148746)
            (1.0, -0.00861864909529686)
            (1.0, -0.013218495063483715)
            (1.0, -0.007036347407847643)
            (1.0, -0.006690574809908867)
            (1.0, -0.009372889064252377)
            (1.0, -0.009229639545083046)
            (1.0, -0.010430139489471912)
        };

    \addplot+[boxplotcolor=black, mark=*,boxplot prepared={
        median=0.0025580484652891755,
        lower quartile=-0.0011100804549641907,
        upper quartile=0.00592587748542428,
        lower whisker=-0.01117649581283331,
        upper whisker=0.014929325319826603,
        lower notch=0.002115836395355651,
        upper notch=0.0030002605352227}
        ]
        coordinates {
            (1.0, -0.013396102003753185)
            (1.0, -0.013957379385828972)
            (1.0, -0.015022761188447475)
            (1.0, -0.013720267452299595)
            (1.0, -0.012393737211823463)
            (1.0, -0.01244555227458477)
            (1.0, 0.017348842695355415)
            (1.0, 0.01850530505180359)
            (1.0, 0.017009630799293518)
            (1.0, 0.01655145362019539)
            (1.0, 0.01779009774327278)
            (1.0, 0.01909431628882885)
        };
  \end{axis}
\end{tikzpicture}%
      \begin{tikzpicture}
  \begin{axis}[
    xlabel={SNR [\si{\dB}]},
    title={CLC \small($\Delta\text{SNR}_t=100$)},
    xmin=0,
    xmax=6,
    ymin=-0.05,
    ymax=0.17,
    grid=major,
    xtick={1, 2, 3, 4, 5},
    xticklabels={-5, 0, 5, 10, 20},
    yticklabels={},
    ytick={},
    boxplot/draw direction=y,
    width=0.5\linewidth,
    height=0.6\linewidth,
    boxplotcolor/.style={color=#1,mark options={color=#1,fill=white}},
    ]

    \addplot+[boxplotcolor=black, mark=*,boxplot prepared={
        median=0.06931721419095993,
        lower quartile=0.049439869821071625,
        upper quartile=0.08641486242413521,
        lower whisker=-0.00498642073944211,
        upper whisker=0.14158904552459717,
        lower notch=0.06699146046779302,
        upper notch=0.07164296791412685}
        ]
        coordinates {
            (1.0, -0.009622856974601746)
            (1.0, -0.01868387684226036)
            (1.0, -0.024788692593574524)
            (1.0, -0.013454105705022812)
            (1.0, -0.25879356265068054)
            (1.0, -0.010027785785496235)
            (1.0, -0.03195849433541298)
            (1.0, -0.01997685432434082)
            (1.0, -0.009973651729524136)
            (1.0, -0.0419175922870636)
            (1.0, 0.16402499377727509)
            (1.0, 0.14383983612060547)
            (1.0, 0.14231158792972565)
            (1.0, 0.26439395546913147)
        };

    \addplot+[boxplotcolor=black, mark=*,boxplot prepared={
        median=0.0747639462351799,
        lower quartile=0.05891764163970947,
        upper quartile=0.08967120572924614,
        lower whisker=0.013163547962903976,
        upper whisker=0.1282508820295334,
        lower notch=0.07283107549475071,
        upper notch=0.0766968169756091}
        ]
        coordinates {
            (1.0, -0.006582931615412235)
            (1.0, -0.006963267922401428)
            (1.0, -0.0015872834483161569)
            (1.0, -0.007800720166414976)
            (1.0, 1.5733703548903577e-05)
            (1.0, 0.005493918899446726)
            (1.0, -0.0013221156550571322)
            (1.0, -0.0160481259226799)
            (1.0, 0.006444680504500866)
            (1.0, -0.009264564141631126)
            (1.0, 0.0023562898859381676)
            (1.0, -0.007820944301784039)
            (1.0, -0.001650015590712428)
            (1.0, 0.0009969836100935936)
            (1.0, 0.010698574595153332)
            (1.0, -0.0007999945082701743)
            (1.0, 0.005985204130411148)
            (1.0, -0.004702379461377859)
            (1.0, -0.000954188930336386)
            (1.0, -0.0038533550687134266)
            (1.0, 0.0008586719050072134)
            (1.0, -0.030834872275590897)
            (1.0, -0.00010948570707114413)
            (1.0, 0.011367462575435638)
            (1.0, -0.011402590200304985)
            (1.0, 0.14710554480552673)
            (1.0, 0.1479194164276123)
            (1.0, 0.13903582096099854)
            (1.0, 0.14205524325370789)
            (1.0, 0.24376994371414185)
        };

    \addplot+[boxplotcolor=black, mark=*,boxplot prepared={
        median=0.046741705387830734,
        lower quartile=0.03255211189389229,
        upper quartile=0.059562708251178265,
        lower whisker=-0.007251414470374584,
        upper whisker=0.09315044432878494,
        lower notch=0.04504408129364309,
        upper notch=0.04843932948201838}
        ]
        coordinates {
            (1.0, -0.009580305777490139)
            (1.0, -0.010006915777921677)
            (1.0, -0.021017882972955704)
            (1.0, -0.018417268991470337)
            (1.0, -0.008985986933112144)
            (1.0, -0.008248960599303246)
            (1.0, -0.016687581315636635)
            (1.0, -0.02107888087630272)
            (1.0, -0.00928187184035778)
        };

    \addplot+[boxplotcolor=black, mark=*,boxplot prepared={
        median=0.02076777908951044,
        lower quartile=0.009809456067159772,
        upper quartile=0.0303593585267663,
        lower whisker=-0.02101031504571438,
        upper whisker=0.05673031136393547,
        lower notch=0.019476211547380863,
        upper notch=0.02205934663164002}
        ]
        coordinates {
            (1.0, -0.027883019298315048)
            (1.0, -0.02467961795628071)
            (1.0, -0.030118096619844437)
            (1.0, -0.02580837346613407)
            (1.0, -0.023628342896699905)
            (1.0, -0.02468952164053917)
            (1.0, -0.024633685126900673)
        };

    \addplot+[boxplotcolor=black, mark=*,boxplot prepared={
        median=-0.002424473175778985,
        lower quartile=-0.007459420594386756,
        upper quartile=0.002379501413088292,
        lower whisker=-0.022146405652165413,
        upper whisker=0.016675135120749474,
        lower notch=-0.0030428523792510043,
        upper notch=-0.0018060939723069658}
        ]
        coordinates {
            (1.0, -0.023501722142100334)
            (1.0, -0.022480616346001625)
            (1.0, -0.030834635719656944)
            (1.0, -0.022612886503338814)
            (1.0, -0.022754520177841187)
            (1.0, -0.024236800149083138)
        };

  \end{axis}
\end{tikzpicture}%
    }
    \vspace{-2em}
    \caption{Comparison with WF-approach \cite{aubreville2018deep}}
    \label{fig:boxplots}
    \vspace{-1.0em}
  \end{figure}

\begin{figure}[htb]
    \vspace{-0.5em}
    \centering
    \includegraphics[width=\columnwidth, trim=9 20 9 5, clip]{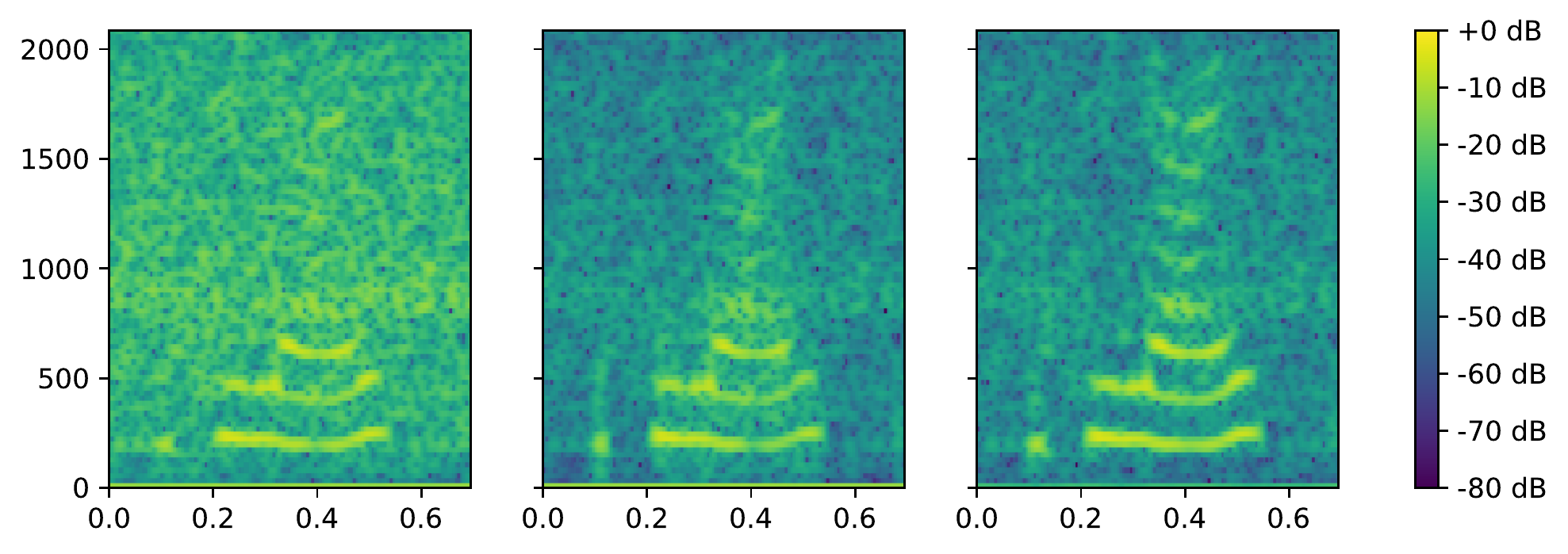}
    \centerline{(a) \hspace{1.9cm} (b) \hspace{1.9cm} (c) \hspace{0.4cm}}
    \vspace{-1.5em}
    \caption{Detail view of \Fig \ref{fig:melspectrograms} (\SIrange{7.5}{8}{\s}). Noisy input (a),
    WF \cite{aubreville2018deep} (b), CLC (c). CLC is able to reduce the noise between harmonics
    within a single frequency band, while WF is limited by the frequency band width and thus, can
    only reduce the noise before and after speech segments.}
    \label{fig:detail}
    \vspace{-1.0em}
  \end{figure}

  The strength of complex linear coding is evident for low SNRs like \SIrange{0}{-5}{\dB}. \Fig
  \ref{fig:boxplots} compares the \dSTOI\ metric with the WF approach and shows a significant
  improvement at \SI{-5}{\dB} and \SI{0}{\dB}. Interestingly, limiting the attenuation via
  $\dSNRt$ to \SI{14}{\dB} yields slightly better results and smaller interquartile
  range \wrt \dSTOI. Due to the 5 coefficients per TF bin, CLC is able to reduce only parts of
  one frequency band, reducing noise between harmonics, while preserving most of the speech. \Fig
  \ref{fig:detail} shows this effect in comparison with the conventional Wiener-Filter approach.

  For high SNRs however, the Wiener-Filter often performs satisfactory. Here, noise between the
  harmonics only has little impact on the phase, which would impair the listening experience. The
  WF with its linear mapping between noisy and clean filter bank representation is perfectly suited
  for these cases. While CLC with $l>=0$ could also learn this, it is a lot harder since it needs to
  zero almost all coefficients and only keep the real part of the coefficient of the current frame.
  Furthermore, the attenuation of a WF can be modified before being applied to the signal, so a
  model can be trained without ``noisy'' input via $\dSNRt$. This allows the network
  to easily remove all noise for high SNR conditions.

  \vfil

\section{Conclusion}
\label{sec:conclusion}

  In this work we presented a real-time capable noise reduction framework for low resolution
  spectrograms based on complex linear coding. We showed that our CLC framework is able to reduce
  noise within individual frequency bands while preserving the speech harmonics. Especially for low
  SNRs, objective metrics show that CLCNet significantly outperforms previous work.

\vfill\pagebreak

\begingroup
\fontsize{9.55pt}{12.0pt}\selectfont
  \bibliographystyle{IEEEbib}
  \bibliography{refs}
\endgroup

\end{document}